\documentclass[10pt,conference]{IEEEtran}

\usepackage[english]{babel} 
\usepackage{amsmath}                
\usepackage{amsfonts}               
\usepackage{amssymb}                
\usepackage{amsopn}                 
\allowdisplaybreaks[4] 
\usepackage{bbm}                    
\usepackage{mathrsfs}               
\usepackage{calc}                   
\usepackage[dvips]{graphicx}        
\usepackage{epsfig}                
\usepackage{psfrag}                 
\usepackage[dvips]{color}           
\usepackage{fancyhdr}               
\usepackage{verbatim}               
\usepackage{exscale}                
\usepackage{research5}
\newtheorem{theorem}{Theorem}
\newtheorem{lemma}[theorem]{Lemma}

\newtheorem{remark}{Remark}
\newtheorem{example}{Example}

\newcommand{\CFB}{C_{\textnormal{FB}}(\SNR)}

\title{A Hot Channel}

\author{\authorblockN{Tobias Koch ~~~ Amos Lapidoth}
\authorblockA{ETH
  Zurich, Switzerland\\
Email: \{tkoch, lapidoth\}@isi.ee.ethz.ch}
\and 
\authorblockN{Paul P.~Sotiriadis}
\authorblockA{Johns Hopkins University, Baltimore, USA\\Email: pps@jhu.edu}}

\begin{document}

\maketitle

\begin{abstract}
This paper studies on-chip communication with non-ideal heat
sinks. A channel model is proposed where
the variance of the additive noise depends on the weighted sum of the
past channel input powers. It is shown that, depending on the weights,
the capacity can be either bounded or unbounded in the input power. A
necessary condition and a sufficient condition for the capacity to be
bounded are presented.
\end{abstract}

\section{Introduction}
\label{sec:intro}
Continuous advancement in VLSI technologies has resulted in extremely small
transistor sizes and highly complex microprocessors. However, on-chip
interconnects responsible for on-chip communication have been improved only
moderately. This leads to the ``paradox'' that local information processing is
done very efficiently, but communicating information between on-chip units
is a major challenge.

This work focuses on an emergent issue expected to challenge circuit
development in future technologies: information communication and processing
is associated with energy dissipation into heat which raises the
temperature of the transmitter/receiver or processing
devices; moreover, the
intrinsic device noise level depends strongly and increasingly on
the temperature. Therefore, the total physical structure can be modeled as a
communication channel whose noise level is data dependent.


This channel was studied at low transmit power levels in
\cite{kochlapidothsotiriadis07_1} where it was shown that in the low
power limit the heating effect is beneficial. In this paper, we focus
on the high transmit power
case. When the allowed transmit power is large, then there is a
trade-off between optimizing the present
transmission and minimizing the interference to future
transmissions. Indeed, increasing the transmission power may help to
overcome the present ambient noise, but it also heats up the chip and
thus increases the noise variance in future receptions. \emph{Prima facie} it
is not clear that, as we increase the allowed transmit power, the
capacity tends to infinity. This paper studies conditions under which
the capacity is bounded in the transmit power.

\subsection{Channel Model}
\label{sub:channelmodel}
\begin{figure}
 \centering
 \psfrag{T}[cc][cc]{Transmitter}
 \psfrag{C}[cc][cc]{Channel}
 \psfrag{R}[cc][cc]{Receiver}
 \psfrag{D}[cc][cc]{Delay}
 \psfrag{M}[b][b]{$M$}
 \psfrag{Mh}[b][b]{$\hat{M}$}
 \psfrag{X}[b][b]{$X_k$}
 \psfrag{Y}[b][b]{$Y_k$}
 \psfrag{Yr}[b][b]{$Y_1^{k-1}$}
 \epsfig{file=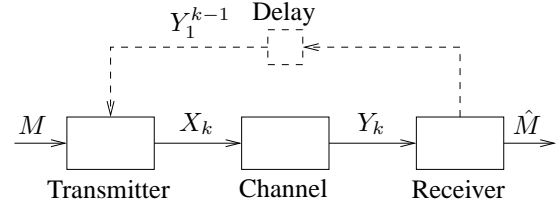, width=0.4\textwidth}
 \caption{The communication system.}
 \label{fig1}
\end{figure}

We consider the communication system depicted in
Figure~\ref{fig1}. The message $M$ to be transmitted over the channel
is assumed to be uniformly distributed over the set
$\set{M}=\{1,\ldots,|\set{M}|\}$ for some positive integer
$|\set{M}|$. The encoder maps the message to the length-$n$ sequence
$X_1,\ldots,X_n$, where $n$ is called the \emph{block-length}. Thus,
in the absence of feedback, the sequence $X_1^n$ is a function of the message
$M$, i.e., $X_1^n=\phi_n(M)$ for some mapping $\phi_n:
\set{M} \to \Reals^n$. Here, $A_m^n$ stands for $A_m,\ldots,A_n$, and
$\Reals$ denotes the set of real numbers.
If there is a feedback link, then $X_k$, $k=1,\ldots,n$ is a function of
the message $M$ and, additionally, of the past channel output symbols
$Y_1^{k-1}$, i.e., $X_k=\varphi_n^{(k)}\big(M,Y_1^{k-1}\big)$ for some mapping
$\varphi_n^{(k)}: \set{M} \times \Reals^{k-1} \to \Reals$.
The receiver guesses the
transmitted message $M$ based on the $n$ channel output symbols
$Y_1^n$, i.e., $\hat{M}=\psi_n(Y_1^n)$ for some mapping $\psi_n: \Reals^n \to \set{M}$.

Let $\Integers^+$ denote the set of positive integers.
The channel output $Y_k \in \Reals$ at time $k\in\Integers^{+}$
corresponding to the channel inputs $(x_1,\ldots,x_k) \in \Reals^k$ is given
by
\begin{equation}
  Y_k = x_k + \sqrt{\left(\sigma^2+\sum_{\ell=1}^{k-1}\alpha_{k-\ell}x_{\ell}^2\right)}\cdot U_k\label{eq:channel}
\end{equation}
where $\{U_k\}$ are independent and identically distributed (IID),
zero-mean, unit-variance
random variables, drawn independently of $M$, and being of finite fourth moment and of
finite differential entropy
\begin{equation}
  h(U_k) > -\infty.\label{eq:finiteentropy}
\end{equation}
The most interesting case is when $\{U_k\}$ are IID, zero-mean,
unit-variance Gaussian random variables, and the reader is encouraged
to focus on this case.
The coefficients $\{\alpha_{\ell}\}$ in \eqref{eq:channel} are non-negative
and bounded, i.e.,
\begin{equation}
  \alpha_{\ell} \geq 0, \quad \ell\in\Integers^+ \quad
  \textnormal{and} \quad \sup_{\ell \in \Integers^+} \alpha_{\ell} < \infty.\label{eq:bounded}
\end{equation}
Note that this channel is not stationary as the variance of the
additive noise depends on the time-index $k$.

We study the above channel under an average-power
constraint on the inputs, i.e.,
\begin{equation}
  \frac{1}{n} \sum_{k=1}^n \E{X_k^2} \leq \const{P}, \label{eq:power}
\end{equation}
and we define the signal-to-noise ratio (SNR) as 
\begin{equation}
  \SNR \triangleq \frac{\const{P}}{\sigma^2}.
\end{equation}

\subsection{Capacity}
\label{sub:capacity}
Let the \emph{rate} $R$ (in nats per channel use) be defined as
\begin{equation}
  R \triangleq \frac{\log |\set{M}|}{n}
\end{equation}
where $\log(\cdot)$ denotes the natural
logarithm function.
A rate is said to be \emph{achievable} if there exists a sequence
of mappings $\phi_n$ (without feedback) or $\varphi_n^{(1)},\ldots,\varphi_n^{(n)}$ (with feedback)
and $\psi_n$ such that the error probability 
$\Prob\big(\hat{M} \neq M\big)$
vanishes as $n$ tends to infinity. The \emph{capacity} $C$ is the
supremum of all achievable rates. We denote by $C(\SNR)$ the capacity
under the input constraint \eqref{eq:power} when there is no feedback, and we add the subscript ``FB''
to indicate that there is a feedback link. Clearly,
\begin{equation}
  C(\SNR) \leq C_{\textnormal{FB}}(\SNR) \label{eq:noFBtoFB}
\end{equation}
as we can always ignore the feedback link.

For the above channel the \emph{capacities per unit cost} which are defined as \cite{verdu90}
\begin{equation}
  \dot{C}(0) \triangleq \sup_{\SNR>0}\frac{C(\SNR)}{\SNR} \quad
  \textnormal{and} \quad \dot{C}_{\textnormal{FB}}(0) \triangleq \sup_{\SNR>0}\frac{C_{\textnormal{FB}}(\SNR)}{\SNR}
\end{equation}
were studied in \cite{kochlapidothsotiriadis07_1} under the additional
assumptions that $\{U_k\}$ are IID, zero-mean, unit-variance Gaussian
random variables, and that the coefficients
fulfill
\begin{equation}
  \sum_{\ell=1}^{\infty} \alpha_{\ell} \triangleq \alpha
  < \infty. \label{eq:alphadef}
\end{equation}
It was shown that, irrespective of
whether feedback is available or
not, the capacity per unit cost is given by
\begin{equation}
  \dot{C}_{\textnormal{FB}}(0) = \dot{C}(0)= \lim_{\SNR \downarrow 0} \frac{C(\SNR)}{\SNR} = \frac{1}{2}(1+\alpha).
\end{equation}

In this paper, we focus on the high SNR case. Specifically, we
explore the question whether the capacity is bounded or unbounded in the
SNR.

\subsection{The Main Result}
\label{sub:result}
We show that whether the capacity is bounded or not depends highly on the decay rate of
the coefficients $\{\alpha_{\ell}\}$.
This is stated precisely in the
following theorem.

\begin{theorem}
  Consider the above channel model. Then,
  \begin{IEEEeqnarray}{llCl}
    \textnormal{i)} \quad & \varliminf_{\ell \to \infty}
    \frac{\alpha_{\ell+1}}{\alpha_{\ell}} > 0 \quad &\Longrightarrow &
    \quad  \sup_{\SNR > 0} C_{\textnormal{FB}}(\SNR)
    < \infty \IEEEeqnarraynumspace \label{eq:main1}\\
    \textnormal{ii)} & \varlimsup_{\ell \to \infty}
    \frac{\alpha_{\ell+1}}{\alpha_{\ell}} = 0 \quad & \Longrightarrow
    & \quad  \sup_{\SNR > 0} C(\SNR) = \infty \label{eq:main2}
  \end{IEEEeqnarray}
  where we define, for any $a>0$, $a/0 \triangleq \infty$ and $0/0\triangleq 0$.
  \label{thm:main}
\end{theorem}

For example, when $\{\alpha_{\ell}\}$ is a geometric sequence, i.e.,
\mbox{$\alpha_{\ell} = \rho^{\ell}$} for $0<\rho<1$, then the capacity is bounded.

\begin{remark}
Part i) of Theorem~\ref{thm:main} holds also when $U_k$ has an
infinite fourth moment. In Part ii) of Theorem~\ref{thm:main}, the
condition on the left-hand side (LHS) of \eqref{eq:main2} can be
replaced by the weaker condition
\begin{equation}
  \lim_{\ell \to \infty}
  \frac{1}{\ell}\log\frac{1}{\alpha_{\ell}} = \infty.
\end{equation}
\end{remark}

A proof of Theorem~\ref{thm:main} is given in the next section. In
Section~\ref{sec:discussion} we address the case where neither the LHS
of \eqref{eq:main1} nor the LHS of \eqref{eq:main2} holds, i.e.,
$\varlimsup_{\ell \to \infty} \alpha_{\ell+1}/\alpha_{\ell} > 0$ and
$\varliminf_{\ell \to \infty} \alpha_{\ell+1}/\alpha_{\ell} = 0$. We
show that in this case the capacity can be bounded or unbounded.

\section{Proof of Theorem~\ref{thm:main}}
\label{sec:proofthm}
In this section we provide a proof of Theorem~\ref{thm:main}. Part
i) is proven in the next subsection, while the proof of Part ii)
can be found in the subsequent subsection.

\subsection{Bounded Capacity}
\label{sub:bounded}
In order to show that
\begin{equation}\varliminf_{\ell \to
  \infty} \frac{\alpha_{\ell+1}}{\alpha_{\ell}}>0 \label{eq:condi}
\end{equation}
implies that the feedback capacity
$\CFB$ is bounded, we derive a capacity upper bound which is, like in
\cite[Sec.~8.12]{coverthomas91}, based on Fano's inequality and on an
upper bound on $\frac{1}{n}I(M;Y_1^n)$. To simplify notation, we
define $\alpha_0\triangleq 1$.

We first note that, due to
\eqref{eq:condi}, we can find an $\ell_0 \in \Integers^+$ and a $0<\rho<1$ so
that 
\begin{equation}
  \alpha_{\ell_0} > 0 \qquad \textnormal{and} \qquad \frac{\alpha_{\ell+1}}{\alpha_{\ell}} \geq \rho, \quad \ell  \geq
  \ell_0. \label{eq:rho}
\end{equation}
We continue with the chain rule for mutual information
\cite{coverthomas91}
\begin{IEEEeqnarray}{lCl}
  \IEEEeqnarraymulticol{3}{l}{\frac{1}{n} I(M;Y_1^n)}\nonumber\\
  & = & \frac{1}{n} \sum_{k=1}^{\ell_0}
  I\big(M;Y_k\big|Y_1^{k-1}\big)+\frac{1}{n}\sum_{k=\ell_0+1}^n
  I\big(M;Y_k\big|Y_1^{k-1}\big).\IEEEeqnarraynumspace \label{eq:firstsum}
\end{IEEEeqnarray}
Each term in the first sum on the right-hand side (RHS) of
\eqref{eq:firstsum} is upper-bounded by
\begin{IEEEeqnarray}{lCl}
  \IEEEeqnarraymulticol{3}{l}{I\big(M;Y_k\big|Y_1^{k-1}\big)} \nonumber\\
  & \leq & h(Y_k) - h\big(Y_k\big|Y_1^{k-1},M\big) \nonumber\\
  & = & h(Y_k)-\frac{1}{2}\E{\log\left(\sigma^2+\sum_{\ell=1}^{k-1}
      \alpha_{k-\ell}X_{\ell}^2\right)} - h(U_k)\nonumber\\
  & \leq & \frac{1}{2}\log\left(2\pi e\left(1+\sum_{\ell=1}^k \alpha_{k-\ell}
    \frac{\E{X_{\ell}^2}}{\sigma^2}\right)\!\right) - h(U_k)
  \nonumber\\
  & \leq & \frac{1}{2}\log\left(2\pi e\left(1+ \big(\sup_{\ell' \in \Integers^+_0}
  \alpha_{\ell'}\big) \cdot \sum_{\ell=1}^k
    \frac{\E{X_{\ell}^2}}{\sigma^2}\right)\!\right) - h(U_k)
  \nonumber\\
  & \leq & \frac{1}{2}\log\left(2\pi e\left(1+\big(\sup_{\ell' \in \Integers^+_0}
  \alpha_{\ell'}\big)\cdot n \cdot\SNR\right)\!\right)-h(U_k)\IEEEeqnarraynumspace\label{eq:firstterm}
\end{IEEEeqnarray}
where $\Integers^+_0$ denotes the set of
non-negative integers. Recall that $\sup_{\ell' \in \Integers^+_0}
\alpha_{\ell'}$ is assumed to be finite. Here, the first inequality follows because conditioning cannot
increase entropy; the following equality follows because $X_1^k$ is
a function of $\big(M,Y_1^{k-1}\big)$ and from the behavior of entropy under
translation and scaling \cite[Thms.~9.6.3 \& 9.6.4]{coverthomas91}
in conjunction with the fact that $U_k$ is independent of $\big(X_1^k,M,Y_1^{k-1}\big)$; the subsequent inequality follows
from the entropy maximizing property of Gaussian random variables
\cite[Thm.~9.6.5]{coverthomas91} and by lower-bounding \mbox{$\E{\log\left(\sigma^2+\sum_{\ell=1}^{k-1}
      \alpha_{k-\ell}X_{\ell}^2\right)} \geq \log\sigma^2$};
the next inequality by upper-bounding each coefficient
\mbox{$\alpha_{\ell} \leq \sup_{\ell' \in \Integers_0^+} \alpha_{\ell'}$}, $\ell=1,\ldots,k$;
and the last inequality follows from $\sum_{\ell=1}^k
\E{X_{\ell}^2}/\sigma^2 \leq n\cdot \SNR$, $k \leq n$
which is a consequence of the power constraint \eqref{eq:power} and of the fact that
$\E{X_{\ell}^2}/\sigma^2 \geq 0$, $\ell \in \Integers^+$.

The terms in the second sum on the RHS of
\eqref{eq:firstsum} are upper-bounded using the general upper bound
for mutual information \cite[Thm.~5.1]{lapidothmoser03_3}
\begin{equation}
  I(X;Y) \leq \int D\big(W(\cdot|x) \big\| R(\cdot)\big) \d Q(x) \label{eq:duality}
\end{equation}
where $D(\cdot\|\cdot)$ denotes relative entropy, $W(\cdot|\cdot)$
is the channel law, $Q(\cdot)$ is the distribution on the channel
input $X$, and $R(\cdot)$ is any distribution on the output alphabet.
Thus, any choice of output distribution $R(\cdot)$ yields an upper
bound on the mutual information.

For $k=\ell_0+1,\ldots,n$ we upper-bound $I\big(M;Y_k\big|Y_1^{k-1}=y_1^{k-1}\big)$
for a given $Y_1^{k-1}=y_1^{k-1}$
by choosing $R(\cdot)$ to be of a Cauchy distribution whose density
is given by
\begin{equation}
  \frac{\sqrt{\beta}}{\pi}\frac{1}{1+\beta y_k^2},
  \quad y_k \in \Reals \label{eq:cauchy}
\end{equation}
where we choose the scale parameter $\beta$ to be\footnote{When $y_{k-\ell_0}=0$
  then the density of the Cauchy distribution \eqref{eq:cauchy} is
  undefined. However, this event is of zero probability and has
  therefore no impact on the mutual information $I\big(M;Y_k\big|Y_1^{k-1}\big)$.}
$\beta = 1/(\tilde{\beta} y_{k-\ell_0}^2)$ and
\begin{equation}
  \tilde{\beta} =
  \min\left\{\rho^{\ell_0-1}\cdot\frac{\alpha_{\ell_0}}{\displaystyle
  \max_{0 \leq \ell' < \ell_0} \alpha_{\ell}},\alpha_{\ell_0},\rho^{\ell_0}\right\} \label{eq:beta}
\end{equation}
with $0<\rho<1$ and $\ell_0\in\Integers^+$ given by \eqref{eq:rho}. 
Note that \eqref{eq:rho} together with the assumption that the
coefficients $\{\alpha_{\ell}\}$ are bounded implies that 
\begin{equation}
  0<\tilde{\beta}<1 \qquad \textnormal{and} \qquad \tilde{\beta} \alpha_{\ell} \leq \alpha_{\ell+\ell_0}, \quad
  \ell \in \Integers^+_0. \label{eq:betaalpha}
\end{equation}
Applying \eqref{eq:cauchy} to \eqref{eq:duality} yields
\begin{IEEEeqnarray}{lCl}
  \IEEEeqnarraymulticol{3}{l}{I\big(M;Y_k\big|Y_1^{k-1}=y_1^{k-1}\big)}\nonumber\\
  & \leq &
  \E{\log\left(1+\frac{Y_k^2}{\tilde{\beta}y_{k-\ell_0}^2}\right)}
  +\frac{1}{2} \log \big(\tilde{\beta}y_{k-\ell_0}^2\big) + \log
  \pi\nonumber\\
  & & {} - h\big(Y_k\big|M,Y_1^{k-1}=y_1^{k-1}\big),
\end{IEEEeqnarray}
and we thus obtain, averaging over $Y_1^{k-1}$,
\begin{IEEEeqnarray}{lCl}
  \IEEEeqnarraymulticol{3}{l}{I\big(M;Y_k\big|Y_1^{k-1}\big)}\nonumber\\
  & \leq & \log\pi -
  h\big(Y_k\big|Y_1^{k-1},M\big) + \frac{1}{2}
  \E{\log\big(\tilde{\beta}Y_{k-\ell_0}^2\big)}\nonumber\\
  & & {} + \E{\log\big(\tilde{\beta}Y_{k-\ell_0}^2+Y_k^2\big)}-\E{\log\big(Y^2_{k-\ell_0}\big)} -\log\tilde{\beta}.\IEEEeqnarraynumspace \label{eq:UB1}
\end{IEEEeqnarray}

We evaluate the terms on the RHS of \eqref{eq:UB1} individually. We
begin with
\begin{equation}
  h\big(Y_k\big|Y_1^{k-1},M\big) = \frac{1}{2}\E{\log\left(\sigma^2+\sum_{\ell=1}^{k-1}
      \alpha_{k-\ell} X_{\ell}^2\right)} + h(U_k)
  \label{eq:1}
\end{equation}
where we use the same steps as in \eqref{eq:firstterm}.
The next term is upper-bounded by
\begin{IEEEeqnarray}{lCl}
  \IEEEeqnarraymulticol{3}{l}{
    \E{\log\big(\tilde{\beta}Y_{k-\ell_0}^2\big)}}\nonumber\\
  & = &
    \E{\Econd{\log\left(\tilde{\beta}\big(X_{k-\ell_0}+\sigma_{X_1^{k-\ell_0-1}}\cdot
    U_{k-\ell_0}\big)^2\right)}{X_1^{k-\ell_0}}}
    \nonumber\\
    & \leq &
    \E{\log\left(\tilde{\beta}\Econd{\big(X_{k-\ell_0}+\sigma_{X_1^{k-\ell_0-1}}\cdot
    U_{k-\ell_0}\big)^2}{X_1^{k-\ell_0}}\right)}
  \nonumber\\
  & = & \E{\log\left(\tilde{\beta}X_{k-\ell_0}^2 +
    \tilde{\beta}\sigma^2+\tilde{\beta}\sum_{\ell=1}^{k-\ell_0-1}\alpha_{k-\ell_0-\ell}
    X_{\ell}^2\right)} \nonumber\\
& \leq & \E{\log\left(\sigma^2+\sum_{\ell=1}^{k-\ell_0}\alpha_{k-\ell}X_{\ell}^2\right)}\label{eq:2} 
\end{IEEEeqnarray}
where, conditional on $X_1^{k-1}=x_1^{k-1}$,
  \begin{equation} 
  \sigma^2_{x_1^{k-1}} \triangleq \sigma^2+\sum_{\ell=1}^{k-1}\alpha_{k-\ell}x_{\ell}^2.
  \end{equation}
Here, the first inequality follows from Jensen's inequality applied to the
concave function $\log(x)$, $x>0$; and the second inequality follows
from \eqref{eq:betaalpha}. 

Similarly, we use Jensen's inequality along with \eqref{eq:betaalpha} to upper-bound
\begin{IEEEeqnarray}{lCl}
  \IEEEeqnarraymulticol{3}{l}{\E{\log\big(\tilde{\beta}Y_{k-\ell_0}^2+Y_k^2\big)}}
  \nonumber\\
  & \leq &
  \E{\log\left(2\sigma^2+2\sum_{\ell=1}^{k-\ell_0}\alpha_{k-\ell}X_{\ell}^2+\sum_{\ell=k-\ell_0+1}^k
      \alpha_{k-\ell}X_{\ell}^2\right)} \nonumber\\
  & \leq & \log 2 + \E{\log\left(\sigma^2+\sum_{\ell=1}^k \alpha_{k-\ell}X_{\ell}^2\right)}.\label{eq:3}
\end{IEEEeqnarray}

In order to lower-bound $\E{\log\big(Y_{k-\ell_0}^2\big)}$ we
need the following lemma:
\begin{lemma}
  \label{lemma:expectedlog}
  Let $X$ be a random variable of density
  $f_{X}(x)$, $x \in \Reals$. Then, for any $0<\delta\leq 1$ and $0 <
  \eta < 1$ we have
  \begin{equation}
    \sup_{c \in \Reals}
    \E{\log |X+c|^{-1} \cdot I\{|X+c| \leq \delta\}} \leq \eps(\delta,\eta)
    +\frac{1}{\eta} h^-(X)
  \end{equation}
  where $I\{\cdot\}$ denotes the indicator function; $h^-(X)$ is defined as
  \begin{equation}
    h^-(X) \triangleq \int_{\{x\in\Reals: f_X(x)>1\}} f_X(x)\log
    f_X(x) \d x;
  \end{equation}
  and where $\eps(\delta,\eta)>0$ tends to zero as $\delta \downarrow 0$.
\end{lemma}
\begin{proof}
  A proof can be found in \cite[Lemma 6.7]{lapidothmoser03_3}.
\end{proof}
We write the expectation as
\begin{IEEEeqnarray}{lCl}
  \IEEEeqnarraymulticol{3}{l}{\E{\log\big(Y_{k-\ell_0}^2\big)}}\nonumber\\
  & = &
  \E{\Econd{\log\left(X_{k-\ell_0}+\sigma_{X_1^{k-\ell_0-1}}\cdot
  U_{k-\ell_0}\right)^2}{X_1^{k-\ell_0}}}\IEEEeqnarraynumspace
\end{IEEEeqnarray}
and lower-bound the conditional expectation for a given $X_1^{k-\ell_0}=x_1^{k-\ell_0}$ by
\begin{IEEEeqnarray}{lCl}
  \IEEEeqnarraymulticol{3}{l}{\Econd{\log\left(x_{k-\ell_0}+\sigma_{x_1^{k-\ell_0-1}}\cdot
  U_{k-\ell_0}\right)^2}{x_1^{k-\ell_0}}}\nonumber\\
  & = & \log \sigma_{x_1^{k-\ell_0-1}}^2 - 2\cdot
  \Econd{\log\left|\frac{x_{k-\ell_0}}{\sigma_{x_1^{k-\ell_0-1}}}+U_{k-\ell_0}\right|^{-1}}{x_1^{k-\ell_0}}
  \nonumber\\
  & \geq & \log\sigma_{x_1^{k-\ell_0-1}}^2 -2\eps(\delta,\eta) -\frac{2}{\eta}
  h^-(U_k)+
  \log\delta^2  \label{eq:bla1}
\end{IEEEeqnarray}
for some $0<\delta\leq 1$ and $0<\eta<1$. Here, the
inequality follows by splitting the conditional
expectation into the two expectations given in \eqref{eq:condexpect}
(on the top of the next page) and by
upper-bounding then the first term on the RHS of \eqref{eq:condexpect}
using Lemma~\ref{lemma:expectedlog} and the second
term by $-\log\delta$.
\begin{figure*}[t]
 \begin{IEEEeqnarray}{lCl}
   \Econd{\log\left|\frac{x_{k-\ell_0}}{\sigma_{x_1^{k-\ell_0-1}}}+U_{k-\ell_0}\right|^{-1}}{x_1^{k-\ell_0}}
   & = &
   \Econd{\log\left|\frac{x_{k-\ell_0}}{\sigma_{x_1^{k-\ell_0-1}}}+U_{k-\ell_0}\right|^{-1}\cdot
     I\left\{\left|\frac{x_{k-\ell_0}}{\sigma_{x_1^{k-\ell_0-1}}}+U_{k-\ell_0}\right|\leq
       \delta\right\}}{x_1^{k-\ell_0}}\nonumber\\ 
   & & {} + \Econd{\log\left|\frac{x_{k-\ell_0}}{\sigma_{x_1^{k-\ell_0-1}}}+U_{k-\ell_0}\right|^{-1}\cdot
     I\left\{\left|\frac{x_{k-\ell_0}}{\sigma_{x_1^{k-\ell_0-1}}}+U_{k-\ell_0}\right| >
       \delta\right\}}{x_1^{k-\ell_0}}  \IEEEeqnarraynumspace\label{eq:condexpect}
\end{IEEEeqnarray}
\rule{\textwidth}{0.5pt}
\end{figure*}
Averaging \eqref{eq:bla1} over $X_1^{k-\ell_0}$ yields
\begin{IEEEeqnarray}{lCl}
  \E{\log\big(Y_{k-\ell_0}^2\big)} & \geq & \E{\log\left(\sigma^2 + \sum_{\ell=1}^{k-\ell_0-1}
      \alpha_{k-\ell_0-\ell}X_{\ell}^2\right)}\nonumber\\
  & & {} - 2\eps(\delta,\eta)-\frac{2}{\eta}h^-(U_k)+\log\delta^2.\IEEEeqnarraynumspace \label{eq:4}
\end{IEEEeqnarray}
Note that, since $U_k$ is of unit variance, \eqref{eq:finiteentropy}
together with \cite[Lemma 6.4]{lapidothmoser03_3} implies that
$h^-(U_k)$ is finite.

Turning back to the upper bound \eqref{eq:UB1} we obtain from \eqref{eq:4}, \eqref{eq:3},
\eqref{eq:2}, and \eqref{eq:1}
\begin{IEEEeqnarray}{lCl}
  \IEEEeqnarraymulticol{3}{l}{I\big(M;Y_k\big|Y_1^{k-1}\big)}\nonumber\\
  & \leq & \log 2 + \E{\log\left(\sigma^2+\sum_{\ell=1}^k
      \alpha_{k-\ell}X_{\ell}^2\right)} + \frac{2}{\eta}
  h^-(U_k)\nonumber\\
  & & {} +2\eps(\delta,\eta)-\log\delta^2-\E{\log\left(\sigma^2 + \sum_{\ell=1}^{k-\ell_0-1}
      \alpha_{k-\ell_0-\ell}X_{\ell}^2\right)} \nonumber\\
  & & {} - \log\tilde{\beta}
  +\frac{1}{2}\E{\log\left(\sigma^2+\sum_{\ell=1}^{k-\ell_0}\alpha_{k-\ell}X_{\ell}^2\right)}
  +\log\pi \nonumber\\
  & & {} -  \frac{1}{2}\E{\log\left(\sigma^2+\sum_{\ell=1}^{k-1}
      \alpha_{k-\ell} X_{\ell}^2\right)} - h(U_k) \nonumber\\
  & \leq & \E{\log\left(\sigma^2+\sum_{\ell=1}^k
      \alpha_{k-\ell}X_{\ell}^2\right)} \nonumber\\
  & & {} - \E{\log\left(\sigma^2 + \sum_{\ell=1}^{k-\ell_0-1}
      \alpha_{k-\ell_0-\ell}X_{\ell}^2\right)} + \const{K} 
\label{eq:UB2}
\end{IEEEeqnarray}
where
\begin{equation}
  \const{K} \triangleq \frac{2}{\eta}h^-(U_k)-h(U_k) +
  2\eps(\delta,\eta) +\log\frac{2\pi}{\tilde{\beta}\delta^2}
\end{equation}
is a finite constant, and where the last inequality in \eqref{eq:UB2}
follows because for any $X_{k-\ell_0+1}^{k-1}=x_{k-\ell_0+1}^{k-1}$ we have $\sum_{\ell=k-\ell_0+1}^{k-1} \alpha_{k-\ell}
x_{\ell}^2 \geq 0$. Note that $\const{K}$ does not depend on $k$ as
$\{U_k\}$ are IID.

Turning back to the evaluation of the second sum on the RHS of \eqref{eq:firstsum} we
use that for any sequences $\{a_k\}$ and $\{b_k\}$
\begin{IEEEeqnarray}{lCl}
  \IEEEeqnarraymulticol{3}{l}{\sum_{k=\ell_0+1}^n
 (a_{k}-b_{k})}\nonumber\\
& = & \sum_{k=n-2\ell_0+1}^n (a_k-b_{k-n+3\ell_0}) +
 \sum_{k=\ell_0+1}^{n-2\ell_0}(a_{k}-b_{k+2\ell_0}).\IEEEeqnarraynumspace
 \label{eq:sums}
\end{IEEEeqnarray}
For $k=n-2\ell_0+1,\ldots,n$ we have
\begin{IEEEeqnarray}{lCl}
  \IEEEeqnarraymulticol{3}{l}{\E{\log\left(\frac{\sigma^2+\sum_{\ell=1}^{k}
        \alpha_{k-\ell}X_{\ell}^2}{\sigma^2+\sum_{\ell=1}^{k-n+2\ell_0-1}
        \alpha_{k-n+2\ell_0-\ell}X_{\ell}^2}\right)}}\nonumber\\
  \qquad \qquad \qquad \qquad & \leq & \log\left(1+\big(\sup_{\ell
        \in \Integers^+_0}\alpha_{\ell}\big) \cdot n \cdot \SNR \right)\IEEEeqnarraynumspace\label{eq:wholesum1}
\end{IEEEeqnarray}
which follows by lower-bounding the denominator by $\sigma^2$, and by using then
Jensen's inequality together with the last two inequalities in \eqref{eq:firstterm}.
Thus, applying \eqref{eq:wholesum1} and \eqref{eq:sums} to
\eqref{eq:UB2} yields
\begin{IEEEeqnarray}{lCl}
  \IEEEeqnarraymulticol{3}{l}{\frac{1}{n}\sum_{\ell=\ell_0+1}^n
    I\big(M;Y_k\big|Y_1^{k-1}\big)} \nonumber\\
  & \leq & \frac{n-\ell_0}{n} \const{K} + \frac{2\ell_0}{n} \log\left(1+\big(\sup_{\ell
      \in \Integers^+_0}\alpha_{\ell}\big) \cdot n \cdot \SNR\right) \nonumber\\
  & & {} + \frac{1}{n}\sum_{k=\ell_0+1}^{n-2\ell_0}\E{\log\left(\frac{\sigma^2+\sum_{\ell=1}^k
      \alpha_{k-\ell}X_{\ell}^2}{\sigma^2 + \sum_{\ell=1}^{k+\ell_0-1}
        \alpha_{k+\ell_0-\ell}X_{\ell}^2}\right)} \nonumber\\
  & \leq &  \frac{n-\ell_0}{n} \const{K} + \frac{2\ell_0}{n} \log\left(1+\big(\sup_{\ell
      \in \Integers^+_0}\alpha_{\ell}\big) \cdot n \cdot \SNR\right) \nonumber\\
  & & {} +
    \frac{1}{n}\sum_{k=\ell_0+1}^{n-2\ell_0}\E{\log\left(\frac{\sigma^2+\sum_{\ell=1}^k
      \alpha_{k+\ell_0-\ell}X_{\ell}^2}{\sigma^2 + \sum_{\ell=1}^{k+\ell_0-1}
        \alpha_{k+\ell_0-\ell}X_{\ell}^2}\right)} \nonumber\\
  & & {}  -
  \frac{n-3\ell_0}{n} \log\tilde{\beta} \nonumber\\
  & \leq & \frac{n-\ell_0}{n} \const{K} + \frac{2\ell_0}{n} \log\left(1+\big(\sup_{\ell
      \in \Integers^+_0}\alpha_{\ell}\big) \cdot n \cdot \SNR\right)
    \nonumber\\
    & & {} - \frac{n-3\ell_0}{n} \log\tilde{\beta} \label{eq:UB3}
\end{IEEEeqnarray}
where the second inequality follows by adding $\log\tilde{\beta}$ to
the expectation and by upper-bounding then $\tilde{\beta} \alpha_{\ell} <
\alpha_{\ell+\ell_0}$, $\ell \in \Integers_0^+$ \eqref{eq:betaalpha}; and the
last inequality follows because for any given
$X_{k+1}^{k+\ell_0-1}=x_{k+1}^{k+\ell_0-1}$ we have
$\sum_{\ell=k+1}^{k+\ell_0-1} \alpha_{k+\ell_0-\ell} x_{\ell}^2 \geq
0$.

Combining \eqref{eq:UB3}, \eqref{eq:firstterm}, and
\eqref{eq:firstsum} we obtain
\begin{IEEEeqnarray}{lCl}
  \IEEEeqnarraymulticol{3}{l}{\frac{1}{n} I(M;Y_1^n)}\nonumber\\
  & \leq & \frac{n-\ell_0}{n}
  \const{K}-\frac{n-3\ell_0}{n}\log\tilde{\beta}
  +\frac{\ell_0}{2n}\log (2\pi e) -
  \frac{\ell_0}{n} h(U_k)\nonumber\\
  & & {} + \frac{\ell_0}{n}\frac{5}{2} \log\left(1+\big(\sup_{\ell
      \in \Integers^+_0}\alpha_{\ell}\big) \cdot n \cdot \SNR\right) \IEEEeqnarraynumspace
\end{IEEEeqnarray}
which converges to $\const{K}-\log\tilde{\beta} < \infty$
as we let $n$ go to infinity. With this, we have shown that
$\varliminf_{\ell \to \infty} \alpha_{\ell+1}/\alpha_{\ell} > 0$
implies that the capacity $\CFB$ is bounded.

\subsection{Unbounded Capacity}
\label{sub:unbounded}
We shall show that 
\begin{equation}
\lim_{\ell \to
  \infty}\frac{1}{\ell}\log\frac{1}{\alpha_{\ell}}=\infty \label{eq:weaker}
\end{equation}
implies that
the capacity $C(\SNR)$ in the absence of feedback is unbounded in
the SNR. Part ii) of Theorem~\ref{thm:main} follows then by noting that
\begin{equation}
  \varlimsup_{\ell \to \infty} \frac{\alpha_{\ell+1}}{\alpha_{\ell}}
  = 0 \quad \Longrightarrow \quad \lim_{\ell \to \infty}
  \frac{1}{\ell} \log\frac{1}{\alpha_{\ell}} = \infty.
\end{equation}

We prove the claim by proposing a coding scheme that achieves
an unbounded rate. We first note that \eqref{eq:weaker}
implies that for any $\varrho>0$ we can find an $\ell_0 \in
\Integers^+$ so that
\begin{equation}
  \alpha_{\ell} < \varrho^{\ell}, \quad \ell \geq \ell_0. \label{eq:positivecoefficients}
\end{equation}

If there exists an $\ell_0 \in
\Integers^+$ so that $\alpha_{\ell}=0$, $\ell \geq \ell_0$, then
we can achieve the (unbounded) rate
\begin{equation}
  R = \frac{1}{2L} \log(1+L\cdot \SNR), \quad L \geq \ell_0
\end{equation}
by a coding scheme where the channel inputs $\{X_{kL+1}\}$ are IID,
zero-mean Gaussian random variables of variance $L\const{P}$, and where the other inputs are
deterministically zero. Indeed, by waiting $L$ time-steps, the chip's
temperature cools down to the ambient one so that the
noise variance is independent of the previous channel inputs and we
can achieve---after appropriate normalization---the capacity of the
additive white Gaussian noise (AWGN) channel \cite{lapidoth96}.

For the more general case \eqref{eq:positivecoefficients} we propose
the following encoding and decoding scheme.
Let $x_1^n(m)$, $m \in \set{M}$ denote the codeword sent out by the transmitter
that corresponds to the message $M=m$. We choose some period $L \geq \ell_0$ and
generate the components
$x_{kL+1}(m)$, $m \in \set{M}$, $k = 0,\ldots,\lfloor n/L
  \rfloor-1$ (where $\lfloor \cdot \rfloor$ denotes the floor
function) independently of each other according
to a zero-mean Gaussian law of variance $\const{P}$. The other components are set to zero.\footnote{It follows from
  the weak law of large numbers that, for any $ m\in \set{M}$, $\frac{1}{n}\sum_{k=1}^n
  x_k^2(m)$ converges to $\const{P}/L$ in probability as $n$
  tends to infinity; this guarantees that the probability that a codeword does not
  satisfy the power constraint \eqref{eq:power} vanishes as $n$ tends
  to infinity.}
 
The receiver uses a \emph{nearest neighbor decoder} in order to guess
 $M$ based on the received sequence of channel outputs $y_1^n$. Thus, it
 computes $ \|\vect{y}-\vect{x}(m')\|^2$ for any $m'\in\set{M}$
 and decides on the message that satisfies
 \begin{equation}
   \hat{M} = \arg \min_{m' \in \set{M}} \|\vect{y}-\vect{x}(m')\|^2
 \end{equation}
where ties are resolved with a fair coin flip. Here, $\|\cdot\|$
denotes the Euclidean norm, and $\vect{y}$ and
$\vect{x}(m')$ denote the respective vectors $\{y_{kL+1}\}_{k=0}^{\lfloor
  n/L\rfloor-1}$ and $\{x_{kL+1}(m')\}_{k=0}^{\lfloor
  n/L\rfloor-1}$.

We are interested in the average probability of error $\Prob\big(\hat{M}
  \neq M\big)$, averaged over
all codewords in the codebook, and averaged over all codebooks. Due to
the symmetry of the codebook construction, the probability of error
corresponding to the $m$-th message $\Prob \big(\hat{M} \neq M \,\big|\, M=m\big)$
does not depend on $m$, and we thus conclude that $\Prob \big(\hat{M}
  \neq M\big) = \Prob\big(\hat{M} \neq M\,\big|\, M=1\big)$. We further note that
\begin{equation}
  \Prob\big(\hat{M} \neq M\,\big|\, M=1\big) \leq \Prob\Bigg(\bigcup_{m'=2}^n
  \|\vect{Y}-\vect{X}(m')\|^2 < \|\vect{Z}\|^2\Bigg) \label{eq:errorprob}
\end{equation}
where $\vect{Z} = \{\sigma_{X(1)_1^{kL}}\cdot U_{kL+1}\}_{k=0}^{\lfloor n/L \rfloor-1}$ which is,
conditional on $M=1$, equal to $\|\vect{Y}-\vect{X}(1)\|^2$.
In order analyze \eqref{eq:errorprob} we need the following lemma.

\begin{lemma}
  \label{lemma:typical}
  Consider the channel described in Section~\ref{sub:channelmodel},
  and assume that the coefficients $\{\alpha_{\ell}\}$ satisfy
  \eqref{eq:weaker}. Further assume that $\{X_{kL+1}\}$
  are IID, zero-mean Gaussian random variables of variance
  $\const{P}$. Let the set $\set{D}_{\eps}$ be defined as
  \begin{IEEEeqnarray}{lCl}
    \set{D}_{\eps} & \triangleq & \Bigg\{(\vect{y},\vect{z})\in
  \Reals^{\lfloor n/L\rfloor}\times \Reals^{\lfloor n/L\rfloor}: \nonumber\\
  & & \quad \left|\frac{1}{\lfloor n/L \rfloor} \|\vect{y}\|^2-(\sigma^2+\const{P}+\alpha^{(L)}\cdot \const{P}) \right| < \eps, \nonumber\\
  & & \quad \left|\frac{1}{\lfloor n/L \rfloor}
    \|\vect{z}\|^2-(\sigma^2+\alpha^{(L)} \cdot \const{P}) \right| <
    \eps \qquad \quad \Bigg\} \IEEEeqnarraynumspace \label{eq:setdef}
  \end{IEEEeqnarray}
  with $\alpha^{(L)}$ being defined as $\alpha^{(L)} \triangleq \sum_{\ell=1}^{\infty} \alpha_{\ell L}$.
  Then, 
  \begin{equation}
    \lim_{n \to \infty} \Prob\big((\vect{Y},\vect{Z}) \in \set{D}_{\eps}\big) = 1
  \end{equation}
  for any $\eps > 0$.
\end{lemma}
\begin{proof}
  First note that, since  $U_k$ has a finite fourth moment, our choice of
  input distribution implies that $\E{(\sigma_{X_1^{k-1}}\cdot U_k)^4}<\infty$. This along with
  \eqref{eq:positivecoefficients} yields that the variances 
  $\Var{\frac{1}{\lfloor n/L \rfloor}\|\vect{Y}\|^2}$ 
  and $\Var{\frac{1}{\lfloor n/L \rfloor}\|\vect{Z}\|^2}$ vanish as
  $n$ tends to infinity. The lemma follows then by computing $\E{\frac{1}{\lfloor n/L
  \rfloor}\|\vect{Y}\|^2}$ and $\E{\frac{1}{\lfloor n/L
  \rfloor}\|\vect{Z}\|^2}$ and by Chebyshev's inequality \cite[Sec.~5.4]{gallager68}.
\end{proof}

In order to upper-bound the RHS of \eqref{eq:errorprob} we proceed
along the lines of \cite{lapidoth96}, \cite{lapidothshamai02}. We have
\begin{IEEEeqnarray}{lCl}
  \IEEEeqnarraymulticol{3}{l}{\Prob\Bigg(\bigcup_{m'=2}^n
  \|\vect{Y}-\vect{X}(m')\|^2 < \|\vect{Z}\|^2\Bigg) \leq \Prob\big((\vect{Y},\vect{Z}) \notin \set{D}_{\eps}\big)} \nonumber\\
  \IEEEeqnarraymulticol{3}{r}{{} \!+\! \int_{\set{D}_{\eps}} \Prob
  \Bigg(\bigcup_{m'=2}^n
  \!\|\vect{y}-\vect{X}(m')\|^2 <
  \|\vect{z}\|^2 \,\Bigg| \,(\vect{y},\vect{z}) \!\Bigg) \!\d
  P(\vect{y},\vect{z}) \!\!\IEEEeqnarraynumspace} \label{eq:typical1}
\end{IEEEeqnarray}
and it follows from Lemma~\ref{lemma:typical} that the first term on
the RHS of \eqref{eq:typical1} vanishes as $n$ tends to infinity.
Note that since the
codewords are independent of each other, conditional on $M=1$, the distribution of
$\vect{X}(m')$, $m'=2,\ldots,|\set{M}|$ does not depend on $(\vect{y},\vect{z})$. We upper-bound the
second term on the RHS of \eqref{eq:typical1} by analyzing $\Prob\big(\|\vect{y}-\vect{X}(m')\|^2 <
  \|\vect{z}\|^2\,\big|\,(\vect{y},\vect{z})\big)$ for each $m'=2,\ldots,|\set{M}|$ and
by applying then the union of events bound.

For $(\vect{y},\vect{z}) \in \set{D}_{\eps}$ and $m'=2,\ldots,|\set{M}|$ we have
\begin{IEEEeqnarray}{lCl}
  \IEEEeqnarraymulticol{3}{l}{\Prob\big(\|\vect{y}-\vect{X}(m')\|^2 <
      \|\vect{z}\|^2\,\big|\,(\vect{y},\vect{z})\big)}\nonumber\\
  & \leq & \exp\Bigg\{-s \lfloor n/L
      \rfloor(\sigma^2+\alpha^{(L)}\cdot \const{P}+\eps) + \frac{s
      \|\vect{y}\|^2}{1-2s\const{P}} \nonumber\\
    & & \qquad \qquad \quad {} - \frac{1}{2} \lfloor n/L \rfloor
      \log(1-2s\const{P})\Bigg\}, \quad s<0 \IEEEeqnarraynumspace\label{eq:typical2}
\end{IEEEeqnarray}
which follows by upper-bounding $\|\vect{z}\|^2$ by $\lfloor
n/L\rfloor(\sigma^2+\alpha^{(L)}\cdot \const{P}+\eps)$ and from the Chernoff bound
\cite[Sec.~5.4]{gallager68}. Using that, for $(\vect{y},\vect{z})\in\set{D}_{\eps}$, $\|\vect{y}\|^2 \geq \lfloor
n/L \rfloor(\sigma^2+\const{P}+\alpha^{(L)}\cdot \const{P}-\eps)$ it
follows from the union of events bound and \eqref{eq:typical2}
that \eqref{eq:typical1} goes to zero as $n$
tends to infinity if for some $s<0$
\begin{IEEEeqnarray}{lCl}
  R & < & \frac{s}{L}(\sigma^2+\alpha^{(L)}\cdot \const{P} +\eps) + \frac{1}{2L}\log(1-2s\const{P})\nonumber\\
  & & {} - \frac{s}{L} \frac{\sigma^2+\const{P}+\alpha^{(L)} \cdot \const{P}-\eps}{1-2s\const{P}}.
\end{IEEEeqnarray}
Thus, choosing $s = - 1/2\cdot 1/(1+\alpha^{(L)}\cdot \const{P})$
yields that any rate below
\begin{IEEEeqnarray}{l}
  -\frac{1}{2L}\frac{\sigma^2+\alpha^{(L)}\cdot \const{P}+\eps}{1+\alpha^{(L)}\cdot \const{P}} + \frac{1}{2L}
  \log\left(1+\frac{\const{P}}{1+\alpha^{(L)}\cdot
      \const{P}}\right)\nonumber\\
  {} +\frac{1}{2L}\frac{\sigma^2+\const{P}+\alpha^{(L)}\cdot \const{P}-\eps}{1+\alpha^{(L)}\cdot \const{P}}\frac{1}{1+\frac{\const{P}}{1+\alpha^{(L)}\cdot
    \const{P}}} \label{eq:ach1}
\end{IEEEeqnarray}
is achievable. As $\const{P}$ tends to infinity this
converges to\footnote{The same rate can also be derived by evaluating $\varliminf_{n \to \infty} \frac{1}{n}I(X_1^n;Y_1^n)$
   for a distribution on the channel inputs under which $\{X_{kL+1}\}$ are IID, zero-mean,
   variance-$L\const{P}$ Gaussian random variables while the other inputs are
   deterministically zero. However, as the channel \eqref{eq:channel}
   is not stationary, it is \emph{prima facie} not clear whether there is
   a coding theorem associated with this quantity.}
\begin{equation}
  \frac{1}{2L}\log\left(1+\frac{1}{\alpha^{(L)}}\right) >
  \frac{1}{2L}\log\frac{1}{\alpha^{(L)}}. \label{eq:achievable}
\end{equation}

It remains to show that given \eqref{eq:positivecoefficients} we can
make $\alpha^{(L)}$ arbitrarily small. Indeed,
\eqref{eq:positivecoefficients} implies that
\begin{equation}
  \alpha^{(L)} = \sum_{\ell=1}^{\infty} \alpha_{\ell L} <
  \sum_{\ell=1}^{\infty} \varrho^{\ell L} = \frac{\varrho^L}{1-\varrho^L} \label{eq:alphaL}
\end{equation}
and \eqref{eq:achievable} can therefore be further lower-bounded by
\begin{equation}
  \frac{1}{2L}\log\left(1-\varrho^L\right) + \frac{1}{2}\log\frac{1}{\varrho}.
\end{equation}
Letting $L$ tend to infinity yields then that we can achieve any rate
below $\frac{1}{2}\log\frac{1}{\varrho}$. As this can be made arbitrarily large by
choosing $\varrho$ sufficiently small, we conclude that
$\lim_{\ell \to \infty}
\frac{1}{\ell} \log\frac{1}{\alpha_{\ell}}=\infty$ implies that the capacity is unbounded.

\section{Beyond Theorem~\ref{thm:main}}
\label{sec:discussion}
Theorem~\ref{thm:main} resolves the question whether capacity is
bounded or unbounded in the SNR when the coefficients satisfy
either $\varliminf_{\ell \to \infty} \alpha_{\ell+1}/\alpha_{\ell}>0$
or $\varlimsup_{\ell \to \infty}
\alpha_{\ell+1}/\alpha_{\ell}=0$.
We next address the case where neither condition holds, i.e.,
\begin{equation}
  \varlimsup_{\ell \to \infty}
\frac{\alpha_{\ell+1}}{\alpha_{\ell}} > 0 \quad \textnormal{and} \quad \varliminf_{\ell \to \infty}
\frac{\alpha_{\ell+1}}{\alpha_{\ell}} = 0. \label{eq:beyond}
\end{equation}
Example~\ref{ex:1} exhibits a sequence $\{\alpha_{\ell}\}$ satisfying
\eqref{eq:beyond} for which the capacity is bounded, and
Example~\ref{ex:2} provides a sequence $\{\alpha_{\ell}\}$ satisfying
\eqref{eq:beyond} for which the capacity is unbounded.

\begin{example}
  \label{ex:1}
  Consider the sequence $\{\alpha_{\ell}\}$ where all
  coefficients with an even index are $1$ and all coefficients with an
  odd index are zero. It satisfies \eqref{eq:beyond} because $\varlimsup_{\ell \to \infty}
  \alpha_{\ell+1}/\alpha_{\ell} = \infty$
  and $\varliminf_{\ell \to \infty} \alpha_{\ell+1}/\alpha_{\ell} =
  0$.
  Thus, at even times, the output $Y_{2k}$, $k\in\Integers^+$ only depends on the ``even''
  inputs $\{X_{2\ell}\}_{\ell=1}^k$, while at odd times, the output $Y_{2k+1}$, $k \in
  \Integers^+_0$ only depends on the ``odd'' inputs
  $\{X_{2\ell+1}\}_{\ell=0}^k$. By proceeding along the lines of the
  proof of Part i) of Theorem~\ref{thm:main} while choosing in
  \eqref{eq:cauchy} $\beta=1/y_{k-2}^2$, it can be shown that the
  capacity of this channel is bounded.
\end{example}

\begin{example}
  \label{ex:2}
  Consider the sequence $\{\alpha_{\ell}\}$ where $\alpha_0=1$, where all coefficients
  with an odd index are $1$, and where all other
  coefficients (whose index is an even positive integer) are
  zero. (Again, we have $\varlimsup_{\ell \to \infty}
  \alpha_{\ell+1}/\alpha_{\ell}=\infty$ and $\varliminf_{\ell \to
  \infty} \alpha_{\ell+1}/\alpha_{\ell}=0$.) Using Gaussian inputs of power
  $2\const{P}$ at even times while setting the inputs to be zero at
  odd times, and measuring the channel outputs only at even times,
  reduces the channel to a memoryless additive noise channel and
  demonstrates the achievability of \cite{lapidoth96}
  \begin{equation*}
     R = \frac{1}{4} \log(1+2\cdot\SNR)
  \end{equation*}
  which is unbounded in the SNR.
\end{example}



\end{document}